\documentstyle{mn}

\def\kpc{{\rm\,kpc}}\def\msun{{\rm\, M}_\odot}
\def\spose#1{\hbox to 0pt{#1\hss}}\def\lta{\mathrel{\spose{\lower 3pt\hbox{$\mathchar"218$}}
     \raise 2.0pt\hbox{$\mathchar"13C$}}}
\def\gta{\mathrel{\spose{\lower 3pt\hbox{$\mathchar"218$}}
     \raise 2.0pt\hbox{$\mathchar"13E$}}}
\def\kms{\,{\rm km\,s}^{-1}}

\newif\ifAMStwofonts
\ifoldfss
  \ifCUPmtlplainloaded \else
    \NewTextAlphabet{textbfit} {cmbxti10} {}
    \NewTextAlphabet{textbfss} {cmssbx10} {}
    \NewMathAlphabet{mathbfit} {cmbxti10} {} 
    \NewMathAlphabet{mathbfss} {cmssbx10} {} 
  \fi
  \ifAMStwofonts
    \ifCUPmtlplainloaded \else
      \NewSymbolFont{upmath} {eurm10}
      \NewSymbolFont{AMSa} {msam10}
      \NewMathSymbol{\upi}     {0}{upmath}{19}
      \NewMathSymbol{\umu}     {0}{upmath}{16}
      \NewMathSymbol{\upartial}{0}{upmath}{40}
      \NewMathSymbol{\leqslant}{3}{AMSa}{36}
      \NewMathSymbol{\geqslant}{3}{AMSa}{3E}

    \fi
  \fi
\fi 

\ifnfssone
  \newmathalphabet{\mathit}
  \addtoversion{normal}{\mathit}{cmr}{m}{it}
  \addtoversion{bold}{\mathit}{cmr}{bx}{it}
  \newmathalphabet{\mathbfit} 
  \addtoversion{normal}{\mathbfit}{cmr}{bx}{it}
  \addtoversion{bold}{\mathbfit}{cmr}{bx}{it}
  \newmathalphabet{\mathbfss} 
  \addtoversion{normal}{\mathbfss}{cmss}{bx}{n}
  \addtoversion{bold}{\mathbfss}{cmss}{bx}{n}
  \ifAMStwofonts
    \ifCUPmtlplainloaded \else
      \UseAMStwoboldmath
      \makeatletter
      \new@mathgroup\upmath@group
      \define@mathgroup\mv@normal\upmath@group{eur}{m}{n}
      \define@mathgroup\mv@bold\upmath@group{eur}{b}{n}
      \edef\UPM{\hexnumber\upmath@group}
      \new@mathgroup\amsa@group
      \define@mathgroup\mv@normal\amsa@group{msa}{m}{n}
      \define@mathgroup\mv@bold\amsa@group{msa}{m}{n}
      \edef\AMSa{\hexnumber\amsa@group}
      \makeatother
      \mathchardef\upi="0\UPM19
      \mathchardef\umu="0\UPM16
      \mathchardef\upartial="0\UPM40
      \mathchardef\leqslant="3\AMSa36
      \mathchardef\geqslant="3\AMSa3E
    \fi
  \fi
\fi 

\ifnfsstwo
  \DeclareMathAlphabet{\mathbfit}{OT1}{cmr}{bx}{it}
  \SetMathAlphabet\mathbfit{bold}{OT1}{cmr}{bx}{it}
  \DeclareMathAlphabet{\mathbfss}{OT1}{cmss}{bx}{n}
  \SetMathAlphabet\mathbfss{bold}{OT1}{cmss}{bx}{n}
  \ifAMStwofonts
    \ifCUPmtlplainloaded \else
      \DeclareSymbolFont{UPM}{U}{eur}{m}{n}
      \SetSymbolFont{UPM}{bold}{U}{eur}{b}{n}
      \DeclareSymbolFont{AMSa}{U}{msa}{m}{n}
      \DeclareMathSymbol{\upi}{0}{UPM}{"19}
      \DeclareMathSymbol{\umu}{0}{UPM}{"16}
      \DeclareMathSymbol{\upartial}{0}{UPM}{"40}
      \DeclareMathSymbol{\leqslant}{3}{AMSa}{"36}
      \DeclareMathSymbol{\geqslant}{3}{AMSa}{"3E}
    \fi
  \fi
\fi 

\ifCUPmtlplainloaded \else
  \ifAMStwofonts \else 
    \def\upi{\pi}
    \def\umu{\mu}
    \def\upartial{\partial}
  \fi
\fi

\begin{document}

\title[Dark Matter Problem in Disk Galaxies]
{The Dark Matter Problem in Disk Galaxies}

\author[J.\ Binney, O. Gerhard and J. Silk]
{James Binney,$^1$ Ortwin Gerhard$^2$ and Joseph Silk$^1$\\
$^1$Physics Department, Oxford University, Keble Road, Oxford,
OX1 3NP\\
$^2$Astronomisches Institut, Universit\"at Basel, Venusstrasse 7, CH-4102
Binningen, Switzerland}

\pubyear{2000}

\maketitle

\begin{abstract}
In the generic CDM cosmogony, dark-matter halos emerge too lumpy and centrally
concentrated to host observed galactic disks. Moreover, disks are predicted
to be smaller than those observed. We argue that the resolution of these
problems may lie with  a combination of the effects of protogalactic disks which  would have had a mass comparable to that of the inner dark halo
and be plausibly non-axisymmetric,
 and of massive galactic winds, which at early times may have
carried off as many baryons as a galaxy now contains.  A host of
observational phenomena, from quasar absorption lines and intracluster gas
through the G-dwarf problem point to the existence of such winds. 
Dynamical interactions will
homogenize and smooth the inner halo, and the observed disk will be the relic of a massive outflow. 
The inner
halo expanded after absorbing energy and angular momentum from the ejected
material.  Observed disks formed at the very end of the galaxy formation
process, after the halo had been reduced to a minor contributor to the
central mass budget and strong radial streaming of the gas had died down.
\end{abstract}


\section{Introduction}

High resolution simulations of galaxy formation, incorporating realistic CDM
initial conditions of dark halo formation, generally confirm the existence
of a universal density (NFW) profile in the outer regions of galaxies \cite
{nfw}. Moreover, some groups are now reporting significant central dark
matter density cusps that are as steep as $\propto r^{-\beta}$ with
$\beta\approx 1.5$.  The existence of even a more modest cusp ($\beta\approx
1$, as in the original NFW result) implies that at the current epoch $L_*$
galaxies have two to three times too much dark matter within a 2 to 2.5 disk
scale lengths \cite{nav}.  This conclusion applies both to the Milky Way,
where the mass of the disk can be dynamically estimated from the motions of
stars near the Sun, and to an ensemble of nearby spirals, for which the
Tully--Fisher relation effectively measures a $M/L$ ratio that can be
compared with values predicted by stellar-synthesis models.  The
Tully-Fisher slope and dispersion are accounted for by the high resolution
simulations, but the normalization is discrepant, by about a factor of 3 in
$M/L$ at given surface brightness, rotation velocity and luminosity
\cite{nav}.

Two further problems encountered with the cold dark matter hypothesis are
(i) that the scale-lengths of disks are predicted to be too small by a
factor $\sim 5$ \cite{ste}, and (ii) an order of magnitude more satellites
are predicted than are observed \cite{moo}. Both of these problems are
closely related to the persistence of substructure in high-resolution N-body
simulations of hierarchical models of dark halo formation.

There are two possible avenues for resolution of these problems.  One
approach is to tinker with the particle physics. One may abandon the idea
that CDM is weakly interacting.  There are CDM  particle candidates
for which annihilation rates are of order the weak rate but for which scattering
crossections are of the order the strong interaction \cite{car,mah}. Such
dissipative CDM may erase both the CDM cusps and clumpiness \cite {spe},
but at the price of introducing an unacceptably spherical  inner core in  massive clusters \cite{jor}. One
may suppress the small-scale power on subgalactic scales, either by invoking
broken scale-invariance \cite {kam} or warm dark matter \cite {som}, in the
hope that the structure of massive dark halos will be modified.

Here we adopt the less radical approach of exploring astrophysical
alternatives.  We accept the fundamental correctness of the CDM picture, and
ask (i) could excess dark matter be ejected from the optical galaxy? and
(ii) why do baryons in galaxies currently have more specific angular
momentum than predicted by the simple CDM picture. We argue that these
questions are connected, and that both may be resolved if galaxies have
first absorbed and then ejected a mass of baryons that is comparable to
their current baryonic masses. An earlier paper argued that baryonic winds can imprint cores within dwarf galaxy dark halos \cite{nef}. Here we propose that
energy and angular momentum surrendered by
the ejected baryons have profoundly modified the dark halo within the
current optical massive galaxy. In this picture most protogalactic
material remained gaseous until the period of mass ejection was
substantially complete -- this conjecture is tenable because we have no
reliable knowledge of either the rate at which, or the efficiency with
which, stars form in a protogalactic environment.

In Section 2 we argue for massive galactic outflows. In Section 3 we ask how
the dark halo was modified as a result of processing the material prior to
ejection. Section 4 is concerned with the implications for the
star-formation rate within a gaseous bar. Section 5 sums up our arguments.

\section{Inflow and outflow}

A primary problem with the conventional picture of galaxy formation is that
in all simulations, the baryons lose much of their angular momentum as they
fall into dark-matter haloes \cite{katzgunn,Weil} rather than conserving it
as semi-analytic models of galaxy formation typically assume
\cite{kaufm,granato}. Consequently, whereas in semi-analytic
galaxy-formation models, the baryons are marginally short of angular
momentum to account for the observed disk sizes, in reality they will fall
short by an order of magnitude.

Current estimates of the acquisition of angular momentum by perturbations in
the expanding universe seem robust, as is the prediction that collapsing
baryons will surrender much of their angular momentum. Hence, we should take
seriously the expectation that protogalaxies early on will contain a
considerable mass of low-angular momentum baryons. What becomes of this
material?  Some of it will have been converted into the galaxy's bulge and
central black hole. However, the mass of low-angular momentum material to be
accounted for is comparable to the mass of the current disk, on account of
the substantial factor by which infalling baryons will have been short of
angular momentum. The bulge and central black hole of the Milky Way, by contrast,
between them contain less mass than the disk by a factor $\sim5$. In
galaxies of later
Hubble type, such as M33, the factor can be substantially greater.

Star formation is always associated with conspicuous outflows, which are
thought to be generically associated with accretion disks. Hence, it is
likely that a significant fraction of a protogalaxy's low angular momentum
baryons are ejected in a wind that is powered by star formation, magnetic
torques and black hole accretion.  Observations of star-burst galaxies such
as M82 lend direct support to this conjecture \cite{axontaylor,dahlem}. We
conclude that the observed disks of galaxies plausibly formed from the higher angular
momentum tail of the conventional distribution. In terms of a
spherically-symmetric infall model, we imagine that the baryons that started
out close to the centre of the protogalaxy were mostly ejected. Galactic
disks are formed from the baryons that were initially confined to the
perifery of the volume from which the final galaxy's dark matter was drawn,
or even came from outside this volume -- the theory of primordial
nucleosynthesis assures us that $\sim90\%$ of all baryons lay outside the
spheres  conventionally containing  galactic dark matter, so there is no
shortage of material to work with.  On account of its large galactocentric
radius, the material to which we are appealing  will initially have had {\em
more\/} angular momentum than the disk into which it was destined to settle.

X-ray observations of early-type galaxies and clusters of galaxies provide
strong support for the conjecture that forming galaxies blow powerful winds.
In clusters of galaxies, the metal-enriched ejecta are directly observed
because they have been trapped by the cluster potential. The wide spread in
the X-ray luminosities of the hot atmospheres of giant elliptical galaxies
has been used to argue persuasively that early winds can escape the
potentials of many galaxies, but not those of the most massive systems,
with the consequence that the ejected gas sometimes falls back into the
visible galaxy and gives rise to a `cooling flow' \cite{ercole}.

Independent arguments point to massive outflows early in galactic evolution.
The narrow dispersion in the colour-magnitude diagrams of cluster
ellipticals, both now and at redshifts $z\sim1$ \cite{franx}, implies that
the galaxies' colours are not heavily contaminated by metal-poor stars.
Early outflows would prevent such contamination \cite{kaufcharlot}, \cite{fers}.
Moreover, bulges and the nuclei of elliptical galaxies are enhanced in
$\alpha$-elements (C, O, Mg) relative to Fe \cite{kuntschner}. This
observation seems to require suppression of star formation from material
that has been enriched in Fe by type Ia supernovae.  It is often assumed
that this suppression is achieved by converting all the protogalactic gas to
stars before many type Ia supernovae have exploded, but it could be also be
achieved by a supernova-driven wind.  Models of the chemical evolution of disks
\cite{pra} similarly yield an acceptably small number of metal-poor stars in
the old disk if a supernova-driven wind carries metal-enriched gas out of
the galaxy.  Finally, the detection of old halo white dwarfs with a
frequency and mass range similar to that inferred for MACHOs from the LMC
microlensing experiments \cite{ibata} will, if spectroscopically confirmed 
\cite{ib},
require a substantial protogalactic outflow phase to eliminate from the
protogalaxy heavy elements that would otherwise pollute stars that formed
later and are observed to have low metalliciticies.  

Galactic outflows will have delivered heavy elements to the intergalactic
medium \cite{lehnert}.  This process not only accounts for the observed
metallicities of intracluster gas \cite{renzini}, but may also be
responsible for the metallicities of the low-density gas that is primarily
detected through Ly$\alpha$ absorption in quasar spectra. It is possible
 at low $z$,
significant enrichment of the IGM and ICM might come from dwarf galaxies,
although the low metallicities of the dwarfs argue against this unless the luminosity function is exceptionally steep.
At the
redshifts $z\gta2$ at which the enriched IGM  material is observed, most of the stars in
the nearby dwarf galaxies will not have formed, so the more luminous galaxies would
have necessarily had to dominate unless a new population of early-forming dwarfs is invoked.  However semi-analytic theory predicts that at
$z\gta2$, most star formation is confined to locations at which luminous
galaxies now reside \cite{Baughetal},  \cite{Benson}.  These locations are far removed from
the low-density gas that is observed to contain heavy elements.  Galactic
winds could be responsible for transporting the heavy elements from the
location of the bulk of star formation, to where they are observed.
Moreover, extended metal-enriched absorption systems might arise from
expanding shells that form in galactic winds in the same way that shells
form around planetary nebulae. 

Thus, many lines of argument suggest that outflows from both spheroids and
disks were common, and therefore that significantly more baryons were
involved in the formation of a given galaxy than it now contains. 

\section{Modification of the halo}

As we have seen, the infalling baryons will have lost much of their angular
momentum.  The lost angular momentum is taken up by the halo. In principle,
acquisition of this angular momentum modifies the halo at all radii, but the
modifications are small where dark matter dominates over baryons, and are
profound only interior to the radius at which $M_{\rm disk}\sim M_{\rm
halo}$. Observationally, we know that the baryons are dominant inside the
solar radius, so we expect the halo profile to be substantially modified
there, precisely as the CDM model seemingly requires \cite{nav}. 

There are threee obvious mechanisms by which gas can lose angular momentum
to the halo. Early on, the halo is expected to be triaxial and its principal
axes will rotate slowly if at all. Gas flowing in such a potential rapidly
loses angular momentum, even if its mass is small compared to the mass of
the local halo \cite{katzgunn}. If gas ever accumulates to the degree that
it contributes a non-negligible fraction of the mass interior to some radius
$r$, two other mechanisms for angular-momentum loss become effective:
massive blobs of gas will lose angular momentum through dynamical friction
\cite{starketal,steinmetz}, and a tumbling gaseous bar will lose angular
momentum through resonant coupling \cite{hern}. These last two processes operate
even if the halo becomes axisymmetric, as it may do where gas contributes
significantly to the overall mass budget.

During the earliest stages of galaxy formation, gas will be far from
centrifugal equilibrium and will flow rapidly inwards. We assume that it
loses energy faster than angular momentum, with the consequence that gas
that started out at a given galactocentric radius will eventually settle to
a (possibly elliptical) ring. If this ring is not substantially
self-gravitating, it will evolve little if at all. Low-surface-brightness
galaxies would seem to be made up of such inert rings of gas.

If the ring is significantly self-gravitating, it will continue to lose
angular momentum to the local halo by a combination of dynamical friction
and bar-driven resonant coupling. The dynamics of a tumbling gaseous bar
embedded in a dark halo of comparable mass has yet to be carefully studied,
but both analytic calculations and simulations show that, in the case of a
stellar bar, resonant coupling is a rapid process: the time-scale of
angular-momentum loss exceeds the bar's dynamical time by a factor of only a
few \cite{WeinbT,debattS}. Consequently, a bar embedded in a dynamically
significant halo will shrink. This shrinkage will
rapidly enhance the mass fraction of baryons because concentration of the
baryons will be accompanied by expansion of the local halo as it takes up
energy and angular momentum shed by the bar. 

These considerations suggest that, if the baryons ever become dynamically
significant, they will go on losing angular momentum to the halo until they
are dominant, and that dominance is achieved by a combination of the baryons
moving in and the dark matter moving out.  Moreover, chemical evolution
models of the Milky Way disk require about half of the disk to have formed
via late infall \cite{pra}, which implies an extended phase of baryonic
infall. The source of the baryons is likely to be stripped satellites that
are merging with the Milky Way and become dynamically disrupted.  Late
infall may double the mass of the disk, with the consequence that the final
disk is close to maximal, and the role of dark matter within the solar
circle is negligible.

In phase space, orbits at energies
around the bar's corotation energy will be highly chaotic, and the strong
orbital shear that is characteristic of chaos will tend to erase
substructure within the halo near the corotation energy.

The coupling between baryons and dark matter is a fairly local process,
essentially confined to a factor of 2 either side of the baryons' corotation
radius. The processes we have described for one corotation radius presumably
occurred in sequence at a series of radii that increased from very small
values out to scales characteristic of present-day disk galaxies.  If the
arguments of the preceeding section are correct, the dark matter at any
given radius $r$ will interact locally with many different parcels of
baryons during the formation process, as these parcels move through radius
$r$ on their way to the galactic centre and probable ejection from the galaxy.

\section{Bars and star formation}

Since the stars of the current disk are now on nearly circular orbits, they
cannot have formed until after any tumbling gaseous bar had dissolved.  Is
it reasonable to have a bar without significant star formation? The dwarf
galaxy NGC 2915 \cite{bur} is an example of a dark-matter dominated galaxy
with a very extended HI disk revealing a central bar and spiral structure
extending well beyond the optical component. Evidently the Toomre $Q$ of
this system satisfies 
 $$Q_{\rm global} > Q > Q_{\rm local},$$
 where $Q_{\rm global} $ and $Q_{\rm local}$ are the critical values of the
disk instability parameter for global non-axisymmetric and local
axisymmetric instabilities, respectively. One can readily imagine that as
the disk forms, the gas surface density increases and the gas velocity
dispersion drops, so that $Q$ decreases, and the local $Q$ criterion is
subsequently satisfied. In the solar neighbourhood the disk satisfies
 $$Q_{\rm local} \approx \left( {\sigma_g \over 10 {\rm  \ kms}^{-1}} \right)
\left( {15 M_{\odot}
{\rm pc}^{-2} \over \mu_g} \right)$$
 and is marginally unstable. The gas disk of the Milky Way presently
contains about $6\times 10^9\,$M$_{\odot}$. In the transient bar phase, the
effective $Q$ is increased by the ratio of bar streaming velocity to gas
velocity dispersion $\sigma_g$, which amounts to a factor of $\sim 10$.
Hence a gas mass of up to $\sim 10^{11}\,$M$_{\odot}$ can be stabilized
against star formation during the transient bar phase. 

It is clear that  high resolution numerical simulations are required to model the
coupling between the non-axisymmetric protodisk and the dark halo.  These
simulations need to include the effects of baryonic dissipation and
star formation. There may be possible stellar relics of an early massive
bar, that would be recognizable as a disk component of old stars with
significant orbital eccentricity.

\section{Conclusions}

Two serious problems currently plague the CDM
theory of galaxy formation: an excess of dark matter within the optical
bodies of galaxies, and disks that are too small. The second problem
reflects the low angular momentum of infalling matter, and is made worse
when one accepts that infalling baryons will surrender much of their angular
momentum to the dark halo. In consequence, galaxies start with more
low-angular-momentum baryons than they currently hold in their bulges and
central black holes. We have argued that the surplus material was early on
ejected as a massive wind. Many direct and indirect observational arguments
point to the existence of such winds.

Although the angular momentum of the first baryons to fall in was inadequate
for the formation of the disk, it was not entirely negligible, and caused
the inner halo to expand when the latter absorbed it. Similarly, the angular
momentum of the baryons that are now in the disk was originally larger than
it now is, and the surplus angular momentum further expanded the inner halo.
In short, through relieving perhaps twice the baryonic mass of the current
galaxy of angular momentum and energy, the dark halo has become
substantially less centrally concentrated than it was originally, and it
now contributes only a small fraction of the mass within the visible galaxy.
During this refashioning of its inner parts, substructure is likely to have
been erased, leaving the final inner halo smoother both locally and
globally. 

This picture requires the baryonic mass to remain gaseous until the dark
halo has been reduced to a minor contributor to the central mass, and a disk
has formed in which most material is on circular orbits. This conjecture is
plausible for two reasons: (i) the dark halo will be unresponsive to the
collective modes of a gaseous disk, so the disk will not have growing modes
until it dominates the gravitational potential in which it sits, and (ii)
the enhanced orbital shear that is characteristic of closed orbits in a barred
potential cannot be conducive to star formation.  In any case we have
little understanding of what controls the rate of star formation in a
protogalaxy, and we know from the fragility of disks \cite{toth} that disks
formed at the end of the formation process, after merging had all but ceased
and the largest substructure had been erased from bulge and inner halo.

Existing numerical simulations of the interactions of baryons and dark
matter during galaxy formation (e.g., 
Navarro \& Steinmetz, 2000; Benson et al., 2000) 
lack both the mass resolution and some of the physics that is
required to realise the essential ideas employed here. For example, in the
simulations of Benson et al.\ the gravitational softening length is
$10h^{-1}\kpc$, and the basic baryonic resolution element has mass
$\sim4\times10^{10}\msun$ and spurious discreteness effects will be present
on mass scales several times larger. Such simulations neglect magnetic
fields (which are believed to drive winds off accretion disks) and energy
input by both supernovae and the central massive object.

In summary, a considerable mass of
low-angular momentum baryons must have been ejected.  This prediction is a
priori plausible, given observations of winds from star-burst galaxies and
outflows from Lyman-break galaxies, and given the prevalence of outflows in
star-formation regions. The heavy element abundances of hot gas in clusters
of galaxies and in cool, low-density gas observed at redshifts $z\sim2$
through quasar absorption lines are likely to arise through the mixing of
metal-rich ejecta with primordial gas.
The low-angular-momentum material having been ejected, the current disks
formed from the higher-angular-momentum baryons that fell in later.
Since the ejection stage commences only once $M_{\rm baryon}\sim M_{\rm dm}$,
in order for self-gravity to drive gas flows and the ensuing winds,
thereby  causing  the dark-matter distribution to expand and the baryons to
further contract, the visible galaxy is inevitably baryon-dominated but has a
circular speed that is required, via the baryonic mass-loss and
the protogalactic dynamical coupling, to approximately match that of the embedding halo.  Thus the so-called 
`disk--halo conspiracy' \cite {bah} is not really a coincidence but a
consequence of dynamical evolution.

\end{document}